\documentstyle[11pt,newpasp,twoside,epsf]{article}
\def\edcomment#1{\iffalse\marginpar{\raggedright\sl#1\/}\else\relax\fi}
\reversemarginpar

\begin{document}  
\title{Gamma Ray Bursts: open problems} 
\author{Gabriele Ghisellini} 
\affil{Osservatorio Astron. di Brera, via Bianchi 46, I--23807 Merate, Italy}

\begin{abstract} 
The internal/external synchrotron shock scenario has proved very successful
in interpreting the key observations about gamma--ray bursts. 
There still remains, however, some big uncertainties. 
The hottest issue concerns the nature of the progenitor,
but there are also other problems concerning the global
energetics, coupled with the issue of the degree of the collimation 
of the fireball.
To be efficient, internal shocks within the relativistic wind 
must occur with large contrasts of their bulk Lorentz factors, 
and it is not clear yet the role of the Compton drag process in limiting
the velocity differences.
The fireball itself can be ``hot" or ``cold" according to what
accelerates it to ultrarelativistic bulk speeds.
In this respect the recent observations of a black body shape of 
the early phases of a few bursts shed new light on this issue.
The most popular radiation process thought to explain the prompt
emission is synchrotron, but it faces severe problems when
comparing the expected spectrum with observations.
Alternatives are called for.
Emission features in the X--ray afterglow and absorption 
features in the prompt spectra are a powerful diagnostical tool.
Besides shedding light on the nature of the progenitor, they can
constrain the total energy release in a beaming--independent way.

\end{abstract}


\section{Are internal shocks efficient enough?} 

The light curves of the prompt emission of GRBs are 
characterized by very fast variability, with timescales as short
as milliseconds (the record holder being GRB 920229, with 0.22 ms; 
Shaefer \& Walker 1999). 
Furthermore, the time--width of the pulses does not increase during 
the burst duration (Fenimore, Ramirez--Ruiz \& Wu 1999).
These two observational facts led researchers to propose the
``internal shock" scenario for the prompt emission of GRBs:
the central engine works intermittently, producing a modulated
outflow composed of different parts moving with different Lorentz 
factors $\Gamma$, initially separated by a distance $R_0$.
If a relatively slow shell of material is followed by a faster
one, they collide at a distance of order $R_0 \Gamma^2$.
Taking $R_0$ of the order of a few Schwarzchild radii
(for a 1--10 solar masses black hole) and $\Gamma$
of the order of a few hundreds, the two shells collide at 
a distance of the order of $\sim 10^{13}$ cm, just where
(for usual parameters), the shells becomes transparent for
Thomson scattering.
This collision produces a single pulse of ms duration.
The distribution of the time--width of the pulses does not 
change during the total duration of the burst since we have 
the same kind of event repeating itself.

The problem with this scenario is the built--in small efficiency,
since only the relative kinetic energy is available to be dissipated,
and only a fraction of it is given to emitting electrons.
For $\Gamma$--contrasts of order 2, only a few per cent of the
total kinetic energy of the fireballs is liberated.
On the other hand, when the already merged shells are 
slowed down by the circumburst matter by the external shocks, 
a larger fraction of energy can be dissipated.
Why, then, is the afterglow total energy of the same order, and
often smaller than the energy in the prompt emission?

To solve this efficiency problem, Beloborodov (2000) and Kobayashi 
\& Sari (2001) proposed that the distribution of
the $\Gamma$--factors among the different shells is huge,
with $\Gamma$--contrasts up to values of 1000 or more.
But if the Lorentz factor of the shells is distributed
in a large interval, then a very fast shell would move in the
photon field created by the previous collisions, would scatter 
these ambient photons and would produce very high energy 
$\gamma$--rays by the inverse Compton process.
This Compton drag effect can be relevant: it can in fact 
decelerate the fast shells and then it would narrow the range of the 
bulk Lorentz factors of the shells undergoing internal shocks, 
thus lowering their efficiency (Lazzati, Ghisellini \& Celotti 1999).
To illustrate this point, suppose that a pair of shells collide, and
that during the collision their average bulk Lorentz factor is $\Gamma_1$.
Observers with the same Lorentz factor would see the produced radiation 
with an isotropic pattern. Call it the $K^\prime$ frame.
Then assume that a later shell is moving with a much larger bulk Lorentz
factor, $\Gamma_2$. 
In $K^\prime$ this bulk Lorentz factor is of the order of 
$\Gamma^\prime_2\sim \Gamma_2/\Gamma_1$.
Moving in the photon field created by the previous collision, the
later shell boosts a fraction $\tau_{\rm T}/2$ ($\tau_{\rm T}$ is 
the optical depth of the later shell, which at these distances if 
of order 1) of these photons at energies $(\Gamma_2^\prime)^2\gg 1$ 
larger.
The total energy released by the fast shell is therefore 
$\sim \tau_{\rm T} (\Gamma_2^\prime)^2/2$ times larger than what 
released by the previous ones.
As a result of this bulk Compton scattering process, the
fast shell slows down, and its interaction with the shell ahead
will occur with a small $\Gamma$--contrast.
All this is enhanced if the fast shells contains electrons
of already high (random) energies, since the energy boost 
for the scattering process is then larger.
Gruzinov \& M\'esz\'aros (2000) studied this process assuming
that photons can scatter back and forth a few times between
the shells.
However, we note that it is unlikely to have more than one 
scattering event, since during the time needed for the photon 
to go back and forth between the two shells, they have expanded, 
greatly reducing their optical depth.

\section{What is the radiation mechanism of the prompt emission?} 

The most popular emission model proposed for interpreting such
$\gamma$--ray spectra is synchrotron emission by relativistic
electrons in intense magnetic fields (Rees \& M\'esz\'aros  1994;
Katz 1994), but alternative scenarios have been proposed, such as 
Comptonization of low energy photons by thermal or quasi--thermal 
particles (Liang et al. 1997; Ghisellini \& Celotti 1999) or
Compton drag (Lazzati et al. 2000; Ghisellini et al. 2000).  
Furthermore, variants on these basic emission processes have also been
studied, such as jitter radiation (Medvedev 2000),
synchrotron emission from particles with an anisotropic pitch angle
distribution (Lloyd \& Petrosian 2000, 2002); 
thin/thick synchrotron emission from a stratified region (Granot,
Piran \& Sari 2000); synchrotron self--Compton or inverse
Compton off photospheric photons (M\'esz\'aros \& Rees 2000).

Internal shocks accelerate particles and compress magnetic fields,
making the synchrotron process a natural option.
There are however problems with this interpretation, prompting the
alternatives listed above.
The first problem concerns the limiting spectral shape of the low
energy part of the spectrum.
Thin synchrotron emission cannot be harder than $F_\nu \propto \nu^{1/3}$:
this corresponds to see the low frequency emission tail from a monoenergetic
particle distribution, or from a distribution with a low energy cutoff.
However, $\sim$10--15 per cent of bursts show harder spectra 
(e.g. Preece et al. 1998; Ghirlanda Celotti \& Ghisellini 2002).
This led to the concept of ``line of death" of synchrotron emission
as the main radiation process.
An even more serious problem with the synchrotron hypothesis
concerns its efficiency: the cooling time of the particles
emitting the observed energies of a few hundreds keV are
very short, of the order of microseconds.
As a consequence, during any conceivable integration time needed to get 
a spectrum, the particles are cooled down to sub--relativistic energies.
The synchrotron spectrum resulting from a cooling population 
(initially monoenergetic, which remains approximately monoenergetic 
during the cooling, but with an ever decreasing energy)
is $F_\nu \propto \nu^{-1/2}$ (Ghisellini, Celotti \& Lazzati, 2000).  
The vast majority of bursts are harder than that.

The proposed alternatives are not problem--free either: 
\begin{itemize}
\item
synchrotron self absorption requires large (relativistic) electron 
densities, making the inverse Compton process dominant; 
\item 
thermal unsaturated or quasi--saturated Comptonization 
requires intrinsic temperatures of the order of 10 keV or less 
(which correspond to observed spectral peaks a factor $\Gamma$ larger), 
which are somewhat smaller that what derived in steady state, 
pair--equilibrium hot plasmas (Svensson 1984);
\item 
the Compton drag process requires very large seed photon densities, which 
can be ``used only once", i.e. the first shell will scatter all of them 
while it is optically thick for scattering, and later shells (required to 
account for variability) move in an almost photon--free environment;
\item 
synchrotron self--Compton suffers the same problems of the synchrotron process
(i.e. electrons cooling too rapidly) and furthermore the observed peak energy
is more strongly dependent of the electron random energy.
\end{itemize}
Also the variants of the standard synchrotron models have difficulties 
in explaining extremely hard spectra.
One way out is to assume that the spectrum is not dominated by a single radiation
process, but is a combination of a few, which can furthermore change
their relative weight during the burst duration.
While this is somewhat at odds with the idea of internal shocks (i.e. having
the same kind of event occurring many times, with equal basic properties),
this idea may more easily explain why there are trends during the burst
prompt phase (for instance, a hard--to--soft behavior).

\section{Hot or cold fireballs?} 


In the standard hot fireball model there is an initial release 
of energy in a small volume -- i.e. a mini--big bang  --
with the sudden creation of electron--positron pairs,
which soon reach thermal equilibrium (at relativistic 
temperatures) with the trapped radiation 
which accelerates the fireball to relativistic speeds 
(Cavallo \& Rees 1978).
When the temperature of the radiation (as measured in the comoving
frame) drops below $\sim$50 keV the pairs annihilate 
faster than the rate at which they are produced.
Were the fireball made purely by pairs, at this time
it would become transparent, letting the internal radiation escape
to infinity with a black body shape (as after the big--bang).
All energy would escape, and no afterglow can be produced.

The presence of even a small amount of barions, corresponding to 
only $\sim 10^{-6}~ M_\odot$, let the fireball be
opaque to Thomson scattering even after the pair annihilation phase: 
the internal radiation continues to accelerate the fireball until 
most of its initial energy has been converted into bulk motion. 
After this phase the fireball expands at a constant speed and at 
some point becomes transparent and the internal radiation escapes.
The amount of this radiation depends on the mass loading of the fireball
(M\'esz\'aros \& Rees 2000), and it can give a non negligible
contribution to the prompt emission, in addition to the one
produced by internal shocks, especially at early times.
Daigne \& Mochkovitch (2002) have studied this expected black body
component, concluding that the non detection of it poses a problem 
for the hot fireball model.
They then argue that the fireball, besides by its internal pressure,
could be accelerated by Poynting flux.
If the latter acceleration mode is dominating, we have ``cold fireballs".
In this respect, the results of Ghirlanda et al. (2002, see also these 
proceedings), are particularly relevant.
They have analyzed the spectral evolution
of a few extremely hard bursts, and found that their time resolved spectra 
(with integration times of a few tenths of a second) can be very well fitted
by a black body during the first few seconds.
Only later the spectrum becomes non--thermal.
\begin{figure}
\epsfysize=8.5cm 
\hspace{1.5cm}\epsfbox{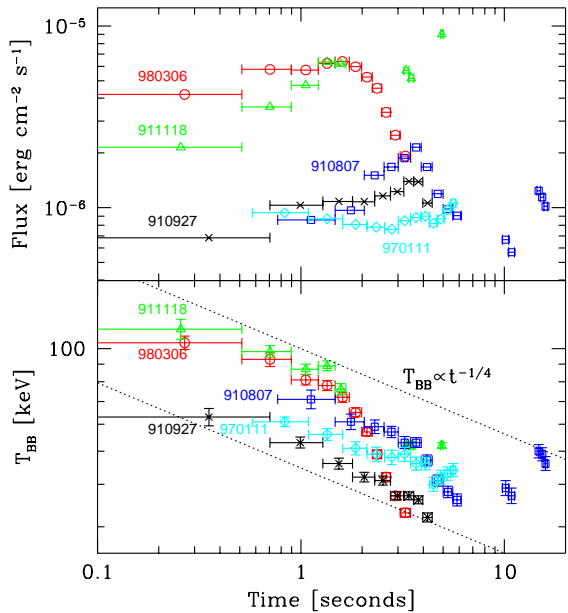} 
\caption{
  Total flux (top panel) and black body temperature (bottom panel) as
  a function of time. The dashed
  lines in the bottom panel 
  correspond to $T_{BB}\propto t^{-1/4}$. From Ghirlanda et al. (2003).}
\end{figure}
Fig. 1 shows the behavior of the flux and the temperature for the bursts
studied by Ghirlanda et al. (2003).
Note that the black body fit to the time resolved spectra are better than a
broken power law fit only for the first few seconds.
Therefore for these bursts there is indeed an initial black body component
in the spectrum, as predicted by the hot fireball scenario.
However, this is not the final proof, because the
fireball could be cold initially, and be heated later 
(but before it becomes transparent).
The heating agent could be magnetic field reconnection
(see e.g. Drenkhahn \& Spruit 2002) or internal shocks occurring early.
The latter case demands that the typical bulk Lorentz factors
of the shells are small enough to let the shells collide before the
transparency radius.

Alternatively, a cold fireball could work if the thermal spectrum we
see is produced by the Compton drag process, boosting circumburst radiation 
to high energies by the fireball bulk motion.

The fact that the black body component is present only initially
can be explained by both scenarios:
if it is photospheric emission, the later black body components compete
with the non--thermal radiation produced in internal shocks and,
in addition, part of black body photons are scattered away from the line of 
sight by the material of the earlier shells;
in the Compton drag scenario, instead, later shells move in a
seed photon--free environment, because most of them have been 
already Compton scattered, and the ``refilling" time is longer than
the duration of the burst.

This suggests a way to discriminate between the two scenarios.
Suppose that the light curve of a burst is characterized by two
peaks, say 10 seconds apart, with a minimum in--between.
In the photospheric model there is the possibility to
have black body radiation also in the initial part of the 
second peak, while this is not expected in the Compton drag scenario.

\section{Strong or mild collimation?} 

If the GRB emission were isotropic, the total energetics 
for the most powerful bursts would reach $10^{54}$ erg, 
a value close to one solar mass entirely converted into energy.
The energy reservoir which can be extracted from a maximally 
spinning 10 solar mass black hole is even greater than that
[being $0.29 Mc^2\sim 5\times 10^{54}M/(10 M_\odot)$ erg],
but it is also reasonable to assume that the emission
is collimated into two oppositely directed cones of
semiaperture angle $\theta_{\rm j}$.
In this case the total energy is decreased, with respect 
to the that calculated assuming isotropy, by the factor
$(1-\cos\theta_{\rm j})\sim \theta_{\rm j}^2/2$.
The total number of bursts is then greater by the same factor,
since we see a GRB only when its jet points at us.
For observers perfectly aligned with the axis of the
fireball cone, one achromatic break should occur
in the lightcurve when $1/\Gamma$ becomes equal
to $\theta_{\rm j}$ (Rhoads 1997; 1999).
Instead, in the much more likely case of a line of sight
making an angle $\theta_0$ with respect to the axis,
there should be {\it two} achromatic breaks:
the first when $1/\Gamma\sim (\theta_{\rm j}-\theta_0)$,
and the second when $1/\Gamma\sim (\theta_{\rm j}+\theta_0)$.
The determination of the jet angle is therefore difficult,
especially when only sparse afterglow data are available,
but despite that Frail et al. (2001) determined the jet angle
for a sample of bursts of known redshifts.
Thus they could correct the ``isotropic energy" of bursts
and obtain the ``true" values.
They found, remarkably, that the corrected values are all very similar
and cluster around a value of a few times $10^{50}$ erg
if the density of the decelerating interstellar medium is
$n=0.1$ cm$^{-3}$.
Assuming larger densities make the jet angle and the ``true" energy
to increase, but the values remain clustered (if the interstellar
density is the same for all bursts). 
Small values of the density are however also in agreement with
the broad band spectral fitting made by e.g. Panaitescu \& Kumar (2001).
The values of the jet angles found in this way range between 
2 and 20 degrees: the average correction factor between 
the isotropic equivalent and the ``true" energy is $\sim 500$.
The found ``true" energy of the bursts is in this case comparable,
or even smaller, than the energy radiated by a supernova.

The problem comes when considering the emission features
seen in a number of X--ray afterglows
(for recent reviews, see B\"ottcher 2002 and Lazzati 2002;
Lazzati, these proceedings).
The energy contained in these emission lines is very large, reaching
$10^{49}$ erg. 
The confidence level of the detections is not huge, being at most
about 3--4$\sigma$ for each case, but they have been observed by 
different detectors on different satellites ({\it Beppo}SAX, ASCA,
and especially XMM--Newton and Chandra).
If real, these lines put a strong and beaming--independent lower 
limit on the total energetic of the bursts, since the efficiency 
to produce an emission X--ray line cannot exceed $\sim$1 per cent at most, 
and therefore their presence requires the burst to have at least 
$\sim10^{51}$ erg in the X--ray band (Ghisellini et al. 2002).
We can reconcile the results of Frail et al. (2001) with
the energetics indicated by the emission lines by increasing the
density of the interstellar medium which makes the calculated
jet opening angles larger, and the energy correction factor smaller.
This however requires a huge increase in the assumed density,
(the calculated opening angle scale as $n^{1/8}$), 
from $n=0.1$ cm$^{-3}$ assumed by Frail et al. (2001) to
$n\sim 10^5$ cm$^{-3}$.

It is furthermore conceivable that the fireballs are not uniform within
$\theta_{\rm j}$, but structured, with an energy per unit solid angle
which is maximum along the axis, and decreasing for increasing angles.
The immediate advantage of this is the fact that the achromatic
jet breaks seen in the afterglow lightcurve would not depend on
$\theta_{\rm j}$, but on $\theta_0$.
In other words, different jet break times would correspond to
different $\theta_0$, not to different $\theta_{\rm j}$:
all bursts could have not only similar total energy, 
but could also be collimated in the same way
(Rossi, Lazzati \& Rees 2002; Zhang \& M\'esz\'aros 2002).

\section{Conclusions}

Since 1997, the beginning of the afterglow era, our knowledge 
of gamma ray bursts has certainly made a quantum jump.
However, problems remains.
Apart from the hunt of the progenitor, clearly a major issue,
the uncertainty in the total energetics is perhaps the
other most important problem to solve.
To know this number we must know the collimation factor
and the efficiency of converting bulk motion into radiation.
The clustering of energies found by Frail et al. (2001) in
this respect is certainly very promising, yet the typical values
found in this way seem to contradict the energetics inferred 
using X--ray emission lines.
An important step forward is then the confirmation of these lines,
and the knowledge of the total energy contained in them, i.e.
their duration.
Knowing the radiation mechanism responsible
for the prompt emission allows us to use it as a diagnostic
tool, to find the physical quantities of the emitting plasma.
In this respect it will be possible, with Swift, to have 
better data (with respect to BATSE) of the low energy part 
(i.e. 10--150 keV) of the prompt spectrum, and hence to put
better constraint to the low energy spectral index.
Few bursts will be pointed by the X--ray telescope aboard Swift
while the burst is still on: for those bursts the frequency 
leverage will be extended down to 0.1 keV, enabling the detection 
of possible absorption features, if they are relatively long lived.

The continuous monitoring in X--rays and in the optical bands 
will allow to search for irregularities in the corresponding
light curves (rebrightenings, microvariabilities, and so on)
which can be used as diagnostic tools to find the distribution of 
the density in the circum--burst medium. 
In this respect the recent experience with GRB 021004 is 
very promising, indicating that when the afterglow can be followed
since very early times (a few minutes after trigger) there are new 
and interesting 
features to be discovered, including the exciting possibility to 
couple the information derived from the irregularities in the
light curve with the spectroscopic information about possible
absorbing material, close to the burst site, in motion with 
slightly different velocities 
(see, among others, Lazzati et al. 2002; Matheson et al. 2003; 
Kobayashi \& Zhang 2002; M\"oller et al. 2002).

Irrespective from any problem we may have to fully understand
the gamma ray burst phenomenon, they are the brightest torchlights
we have to illuminate the far and very far universe; 
furthermore we have a reasonable hope that their existence is
associated to the formation of stars, including the very first stars
beyond the re--ionization epoch.
Since Swift will locate bursts with a few arcminutes error box in a few seconds,
and its optical monitor plus ground based robotic telescopes like REM (see
e.g. Zerbi et al., these proceedings) will give sub--arcsec positions
in a minute, large (infrared) telescope can point objects at redshifts
greater than 10 when the afterglow is still bright enough to allow high 
resolution spectroscopy.

\vskip 0.5 true cm
\noindent
I thank Annalisa Celotti, Stefano Covino, Davide Lazzati, Daniele Malesani,
Elena Rossi and Fabrizio Tavecchio for useful discussions.
I thank the MIUR for funding through COFIN (code 2001022957).


\begin{references}

\reference Beloborodov A.M., 2000, ApJ, 539, L25
\reference B\"ottcher M., 2002, Adv. in Sp. Res., in press (astro--ph/0212034)
\reference Cavallo G. \& Rees M.J., 1978, MNRAS, 183, 359
\reference Daigne F. \& Mochkovitch R., 2002, MNRAS, 336, 1271
\reference Drenkhahn G. \& Spruit H.C., 2002, A\&A, 391, 1141
\reference Fenimore E.E., Ramirez--Ruiz E. \& Wu B., 1999, ApJ, 518, L73
\reference Frail D.A. et al., 2001, ApJ, 562, L55
\reference Ghirlanda G., Celotti A. \& Ghisellini G., 2003, subm. to A\&A
   (astro--ph/0210693)
\reference Ghirlanda G., Celotti A. \& Ghisellini G., 2002, A\&A, 393, 409
\reference Ghisellini G., Lazzati D., Rossi E. \& Rees M.J., 2002, A\&A, 389, L33
\reference Ghisellini G., Celotti A. \& Lazzati D., 2000, MNRAS, 313, L1 
\reference Ghisellini G., Lazzati D., Celotti A. \& Rees M.J., 2000, MNRAS, 316, L45
\reference Ghisellini G. \& Celotti A., 1999, ApJ, 511, L93
\reference Granot J., Piran T. \& Sari R., 2000, ApJ, 534, L163
\reference Gruzinov A. \& M\'esz\'aros P., 2002, ApJ, 539, L24
\reference Katz J.I., 1994, ApJ 432, L107
\reference Kobayashi S. \& Sari R., 2001, ApJ, 551, 934
\reference Lazzati D., Ghisellini G., Celotti A. \& Rees M.J., 2000, ApJ, 529, L17 
\reference Lazzati D., Ghisellini G. \& Celotti A., 1999, MNRAS, 309, L13 
\reference Lazzati D., 2002, NBSI workshop {\it Beaming and Jets in Gamma Ray Bursts}, in press 
     (astro--ph/0211174)
\reference Lazzati D. et al., 2002, A\&A, 396, L5
\reference Liang E. et al., 1997, ApJ, 479, L35
\reference Lloyd N.M. \& Petrosian V., 2000, ApJ, 543, 722
\reference Lloyd N.M. \& Petrosian V., 2002, ApJ, 565, 182
\reference Matheson T., 2003, ApJ, 582, L5
\reference Medvedev M.V., 2000, ApJ, 540, 704
\reference M\'esz\'aros P. \& Rees M.J., 2000, ApJ, 530, 292
\reference M\"oller P. et al., 2002, A\&A, 396, L21
\reference Panaitescu A. \& Kumar P. 2001, ApJ, 554, 667
\reference Preece R.D. et al., 1998, ApJ, 506, L23
\reference Rees M.J. \& M\'esz\'aros P., 1994, ApJ, 430, L93
\reference Rhoads J.E., 1997, ApJ, 487, L1
\reference Rhoads J.E., 1999, ApJ, 525, 737
\reference Rossi E., Lazzati D. \& Rees M.J., 2002, MNRAS, 332, 945
\reference Shaefer B.E. \& Walker K.C., 1999, ApJ. 511, L89
\reference Svensson R., 1984, MNRAS, 209, 175
\reference Zhang B. \& M\'esz\'aros P., 2002, ApJ, 571, 876
\end{references}
\end{document}